\documentstyle[12pt]{article}

\newcommand{\be}{\begin{equation}}
\newcommand{\ee}{\end{equation}}
\newcommand{\bea}{\begin{eqnarray}}
\newcommand{\eea}{\end{eqnarray}}

\newcommand{\dd}{\partial}

\begin{document}
\baselineskip .25in
\newcommand{\numero}{hep-th/9507040, SHEP 95/23}

\newcommand{\titre}{Classical Duality in Gauge Theories}
\newcommand{\auteura}{Noureddine Mohammedi}
\newcommand{\place}{Department of Physics\\University of
Southampton\\
Southampton SO17 1BJ \\ U.K. }
\newcommand{\beq}{\begin{equation}}
\newcommand{\eeq}{\end{equation}}

\newcommand{\abstrait}
{A dual action is obtained for a general non-abelian and
non-supersymmetric gauge theory at the classical level.
The construction of the dual theory follows steps similar
to those used in pure abelian gauge theory. As an example
we study the spontaneously broken $SO(3)$ gauge theory and
show that the electric  and the magnetic fields get
interchanged in the dual theory.}
\begin{titlepage}
\hfill \numero  \\
\vspace{.5in}
\begin{center}
{\large{\bf \titre }}
\bigskip \\ by \bigskip \\ \auteura
\,\,\footnote{e-mail: nouri@hep.phys.soton.ac.uk}
    \bigskip \\ \place \bigskip \\

\vspace{.9 in}
{\bf Abstract}
\end{center}
\abstrait
 \bigskip \\
\end{titlepage}

\newpage
The understanding of electric-magnetic duality is certainly one
one of the challenges of four-dimensional field theories.
This duality was conjuctured by Montonen and Olive \cite{olive}
and shown by Osborn \cite{osborn} to be plausible for
$N=4$ supersymmetric gauge theories.
In fact in the $N=4$ supersymmetric Yang-Mills
case the electric-magnetic duality
was firmly tested by Vafa and Witten
\cite{vafa}.
\par
Recently, dramatic new evidence for the validity of this conjecture
has emerged from the work of A. Sen \cite{sen} and
a version of
Montonen-Olive duality was surprisingly found by Seiberg and
Witten in $N=2$ supersymmetric gauge theory in
four dimensions \cite{seiberg}. This duality has many
similar features with the target space duality encountred in string
theories (see ref.\cite{amit} for a review).
\par
In this paper we examine, at the classical level, the procedure
of constructing a dual theory of a non-abelian and
non-supersymmetric gauge theory.\\
\bigskip\\
{\it The Dual Variables}\\
\bigskip\\
Our starting point is the action for a general gauge theory in
four dimensions. This takes the form
\be
S=\int{\rm d}^4x\sqrt{\gamma}\left[
\gamma^{\mu\rho}\gamma^{\nu\sigma}g_{ab}\left(
\alpha F^a_{\mu\nu} F_{\rho\sigma}^b
+\beta F^a_{\mu\nu}  {\widetilde {F}}_{\rho\sigma}^b \right)+
J^\mu_a(\phi) A_{\mu}^a +
K^{\mu\nu}_{ab}(\phi)A^a_\mu A_\nu^b +S_0(\phi)\right]\,\,\,.
\label{action1}
\ee
Here $\alpha$ and $\beta$ are two coupling constants and
$\phi$ is any generic  matter field on which $J^\mu_a$,
$K^{\mu\nu}_{ab}$ and $S_0$ depend. The field strength is
$F_{\mu\nu}^a=\dd_\mu A_{\nu}^a -\dd_\nu A^a_{\mu}
+\lambda f^a_{bc} A_{\mu}^bA_\nu^c $,
where $\lambda$ is related to $\alpha$, and its dual is
$ {\widetilde {F}}^a_{\mu\nu}=\frac{1}{2}
\epsilon^{\alpha\beta}_{
\,\,\,\,\,\mu\nu}F_{\alpha\beta}^a$.
\par
The action is invariant under the gauge transformations
\be
\delta A^a_\mu=-\dd_\mu \epsilon^a +
\lambda f^a_{bc} A_\mu^c
\epsilon^b
\ee
provided that
\bea
\delta S_0 &=& J^\mu_{a}\dd_\mu\epsilon^a \nonumber\\
\delta J^\mu_a &=& 2K^{\mu\nu}_{ab}\dd_\nu\epsilon^b
+\lambda f^c_{ab}J^\mu_c\epsilon^b
\nonumber\\
\delta K^{\mu\nu}_{ab}&=&+\lambda \left(f^d_{ac}K^{\mu\nu}_{db}
+ f^d_{bc}K^{\mu\nu}_{ad}\right)\epsilon^c\,\,\,\,.
\eea
Now in terms of the self-dual and anti-self-dual projectors
$F^{\pm\,a}_{\mu\nu}=\frac{1}{2}\left(F^a_{\mu\nu} \pm
 {\widetilde {F}}^a_{\mu\nu}\right)$, the action takes the form
\be
S=\int\rm{d}^4x\sqrt{\gamma}\left(
\tau_+ \left(F^+\right)^2 +\tau_- \left( F^-\right)^2
+J^\mu_a A_{\mu}^a +
K^{\mu\nu}_{ab}A^a_\mu A_\nu^b +S_0\right)\,\,\,,
\ee
where $\tau_{\pm}=\alpha \pm \beta $.
\par
Our aim is to find an action which is classically
equivalent to the above action. This we find by introducing
some new degrees of freedom $G^{\pm \,a}_{\mu\nu}$ and the action
we are after is given by
\be
I=\int\rm{d}^4x\sqrt{\gamma}\left(
a\left(G^+\right)^2 +b \left( G^-\right)^2
+c F^+G^+ +d F^-G^-
+J^\mu_a A_{\mu}^a +
K^{\mu\nu}_{ab} A^a_\mu A_\nu^b +S_0\right)\,\,\,.
\label{action2}
\ee
Here $a,b,c$ and  $d$ are some parameters that we are going to
determine. In Maxwell's theory and in the absence of matter fields
one arrives at this action by enlarging the gauge symmetry \cite{witten}.
\par
This action is invariant under two gauge transformations. The first
is given by
\bea
\delta G^{+\,d}_{\mu\nu}&=&
\lambda f^d_{bc}G^{+\,c}_{\mu\nu}
\epsilon^b \nonumber\\
\delta G^{-\,a}_{\mu\nu}&=&
\lambda f^a_{ec}G^{-\,c}_{\mu\nu}
\epsilon^e
\label{gauge1}
\eea
and the second gauge transformation is
\bea
\delta G^{+\,d}_{\mu\nu}&=&
\frac{\lambda c}{2a}f^d_{cb}F^{+\,c}_{\mu\nu}
\epsilon^b \nonumber\\
\delta G^{-\,a}_{\mu\nu}&=&
\frac{\lambda d}{2b}f^a_{ce}F^{-\,c}_{\mu\nu}
\epsilon^e
\label{gauge2}
\eea
\par
To see that this new action is classically equivalent to the
original gauge theory, we eliminate the independent fields
$G^{\pm}$ by their equations of motion
\be
G^+=-\frac{c}{2a} F^+\,\,\,\,,\,\,\,\,
G^-=-\frac{d}{2b} F^-\,\,\,\,.
\label{eqmotion}
\ee
Upon replacing these values for $G^{\pm}$ in the action $I$, we
get back our original action $S$ provided that
\be
\tau_+=-\frac{c^2}{4a}\,\,\,\,\,,\,\,\,\,\,\,
\tau_-=-\frac{d^2}{4b}\,\,\,\,.
\ee
Notice that the two gauge transformations for $G^{\pm}$ are
consistent with the equations of motion for $G^{\pm}$.
\par
The dependence of the action $I$ on the gauge field $A_\mu^a$
is at most quadratic and therefore $A_\mu^a$ can be classically
integrated out. The equation of motion for $A^a_\mu$ is given by
\be
A_\mu^a=-\widetilde R^{ab}_{\mu\nu}V_b^\nu \,\,\,\,,
\label{gaugefield}
\ee
where $\widetilde R^{ab}_{\mu\nu} $ and $V_a^\mu$ are defined
below
\bea
R^{\mu\nu}_{ab} &=& 2K^{\mu\nu}_{ab} + 2\lambda
\gamma^{\mu\alpha}\gamma^{\nu\beta}g_{rs} f^s_{ab}\left(
cG^{+\,r}_{\alpha\beta} +d G^{-\, r}_{\alpha\beta}\right)
\nonumber\\
V^\mu_a&=& J^\mu_a -2\gamma^{\alpha\beta}\gamma^{\mu\nu}
g_{ab}\left(c \dd_\alpha G^{+\,b}_{\beta\nu}+d
\dd_\alpha G^{-\,b}_{\beta\nu}\right)\,\,\,\,
\eea
and $\widetilde R^{ab}_{\mu\nu}$ is the inverse of $R^{\mu\nu}_{ab}$
\be
R^{\mu\nu}_{ab}\widetilde R^{bc}_{\nu\alpha}=\delta^\mu_\alpha
\delta^a_c\,\,\,\,.
\ee
\par
Replacing for $A_\mu^a$ in the action $I$ we obtain the dual action
\be
I_{\rm {dual}}=\int{\rm d}^4x\sqrt{\gamma}\left[
-\frac{c^2}{4\tau_{+}}\left(G^+\right)^2
-\frac{d^2}{4\tau_{-}}\left(G^-\right)^2
-\frac{1}{2}{\widetilde {R}}^{ab}_{\mu\nu} V^\mu_a V^\nu_b
+S_0 \right]\,\,\,\,.
\ee
As expected, the couplings $\tau_{\pm}$ have been inverted.
This action is a generalisation  of a model put forward by
Freedman and Townsend \cite{freedman}
and by Sugamoto \cite{sugamoto}
and which appeared recently in connection
with treating duality using loop space
variables \cite{oxford}.
\par
One can verify that this dual action has inherited the two
gauge invariances (\ref{gauge1}) and (\ref{gauge2}),
up to a total derivative,  where
the fields $F^{\pm}$ are now built from the gauge field $A^a_\mu$ as
given in (\ref{gaugefield}). It is a straightforward calculation
to show that
under the transformation (\ref{gauge1}) we have
\bea
\delta V^\mu_a& = &\lambda f^c_{ab}V_c^\mu \epsilon^b
+ R^{\mu\nu}_{ab}\dd_\nu\epsilon^b \nonumber\\
\delta R^{\mu\nu}_{ab} &=&
\lambda f^d_{ac}R^{\mu\nu}_{db}\epsilon^c
+\lambda f^d_{bc}R^{\mu\nu}_{ad}\epsilon^c    \,\,\,\,.
\eea
This leads to the expetected gauge transformation
for $A^a_\mu$ in (\ref{gaugefield});  $
\delta A^a_\mu=-\dd_\mu \epsilon^a +
\lambda f^a_{bc}A_\mu^c\epsilon^b $.
\par
However, under the gauge transformation
(\ref{gauge2}) we have the
following transformations for $V^\mu_a$ and $R^{\mu\nu}_{ab}$
\bea
\delta V^\mu_a& = &\lambda f^c_{ab}V_c^\mu \epsilon^b
+ R^{\mu\nu}_{ab}\dd_\nu\epsilon^b
-2c\lambda\gamma^{\mu\nu}\gamma^{\alpha\beta}g_{rc}f^r_{ab}
\left(G^{+\,c}_{\nu\alpha}+
\frac{c}{2a}F^{+\,c}_{\nu\alpha}\right)\dd_\beta\epsilon^b
\nonumber\\
&-&2d\lambda\gamma^{\mu\nu}\gamma^{\alpha\beta}g_{rc}f^r_{ab}
\left(G^{-\,c}_{\nu\alpha}+
\frac{d}{2b}F^{-\,c}_{\nu\alpha}\right)\dd_\beta\epsilon^b
+2c\lambda\gamma^{\mu\nu}\gamma^{\alpha\beta}g_{rc}f^c_{ab}
\dd_\alpha\left(G^{+\,r}_{\beta\nu}+
\frac{c}{2a}F^{+\,r}_{\beta\nu}\right)\epsilon^b
\nonumber\\
&+&2d\lambda\gamma^{\mu\nu}\gamma^{\alpha\beta}g_{rc}f^c_{ab}
\dd_\alpha\left(G^{-\,r}_{\beta\nu}+
\frac{d}{2b}F^{-\,r}_{\beta\nu}\right)\epsilon^b
\nonumber\\
\delta R^{\mu\nu}_{ab} &=&
\lambda f^d_{ac}R^{\mu\nu}_{db}\epsilon^c
+\lambda f^d_{bc}R^{\mu\nu}_{ad}\epsilon^c
-2c\lambda^2\gamma^{\mu\alpha}\gamma^{\nu\beta}g_{sc}f^s_{dr}
f^r_{ab}
\left(G^{+\,d}_{\alpha\beta}+
\frac{c}{2a}F^{+\,d}_{\alpha\beta}\right)\epsilon^c
\nonumber\\
&-&2d\lambda^2\gamma^{\mu\alpha}\gamma^{\nu\beta}g_{sc}f^s_{dr}
f^r_{ab}
\left(G^{-\,d}_{\alpha\beta}+
\frac{d}{2b}F^{-\,d}_{\alpha\beta}\right)\epsilon^c
\,\,\,.
\eea
We see that the two transformations coincide when the equations
of motion for $G^{\pm}$ are satisfied. Notice also that varying
$I_{{\rm {dual}}}$ with respect to $G^{\pm}$ leads to the
equations (\ref{eqmotion}) where $A^a_\mu$ is as given by
(\ref{gaugefield}).
\par
It is worth mentioning that the two terms in $I$ involving
the parameters $c$ and $d$ can be written as
\be
cF^+G^+ + dF^-G^-= \frac{1}{2}\left(c+d\right)FG
 +\frac{1}{2}\left(c-d\right)\widetilde F G
\ee
and in the pure abelian gauge theory one usually takes
$c=-d=1$. This choice is crucial in this case because
integrating out the gauge field from $I$ leads to
$\dd_\mu\widetilde G_{\mu\nu}=0$, and by virtue of Poincar\'e's
lemma this in turn implies that
$G_{\mu\nu}=\dd_\mu B_\nu - \dd_\nu B_\mu$, where $B_\mu$ is
an abelian one-form \cite{oxford}.\\
\bigskip\\
{\it The SO(3) Gauge Theory}\\
\bigskip\\
Let us now consider an example and examine the consequences
of this change of variables.
The model we have in mind is the $SO(3)$ gauge theory with a
Higgs triplet.  The corresponding Lagrangian is given by
\be
{\cal L}= -\frac{1}{4}F_{\mu\nu}^aF_{\mu\nu}^a
+\frac{1}{2}\left(D_\mu\phi\right)^a
\left(D_\mu\phi\right)^a
+\frac{\mu^2}{2}\phi^a\phi^a -\frac{\lambda}{4}
\left(\phi^a\phi^a\right)^2\,\,\,,
\ee
where $\mu^2 > 0$, $F_{\mu\nu}^a= \dd_\mu
A_\nu^a-\dd_\nu A_\mu^a + e\epsilon^{abc}A_\mu^b A_\nu^c$
and
$D_\mu\phi^a=\dd_\mu\phi^a +e\epsilon^{abc}A_\mu^b\phi^c$.
The flat space-time metric has the
signature $\left(+,+,+,-\right)$. A quick comparison of this
action  and $S$ in (\ref{action1}) gives the different
quantities used in $S$. In particular $\beta=0$
and $\alpha=-1/4$.
\par
The $SO(3)$ gauge theory has a time-independent solution
of the form
\cite{thooft,polyakov, zee, prasad}
\bea
A^a_i &=&\epsilon_{abi}x_b\left(\frac{K(r)-1}{er^2}\right)
\nonumber\\
A^a_0&=&x_aJ(r)/er^2\nonumber\\
\phi^a&=&x_aH(r)/er^2\,\,\,\,.
\eea
This form solves the equations of motion if the functions
$K$, $J$ and $H$ satisfy the differential equations
\cite{zee}
\bea
r^2J''&=& 2JK^2\nonumber\\
r^2H''&=&2HK^2 -\mu^2r^2 H\left(1-\frac{\lambda}{e^2\mu^2r^2}
H^2\right)\nonumber\\
r^2K''&=&K\left(K^2-J^2+H^2-1\right)\,\,\,.
\eea
\par
After the symmetry breaking one needs to identify the physically
observable fields, especially the photon $F_{\mu\nu}$. A gauge invariant
definition for the electromagnetic field $F_{\mu\nu}$ is given by
\cite{thooft}
\be
{\cal F}_{\mu\nu}=\frac{1}{\phi}\phi^aF_{\mu\nu}^a
-\frac{1}{e\phi^3}\epsilon_{abc}\phi^a
D_\mu\phi^b D_\nu\phi^c\,\,\,,
\ee
where $\phi=\left(\phi^a\phi^a\right)^{1/2}$. This definition
can be cast into the more appealing expression
\be
{\cal F}_{\mu\nu}=\dd_\mu A_\nu -\dd_\nu A_\mu
-\frac{1}{e\phi^3}\epsilon_{abc}\phi^a
\dd_\mu\phi^b \dd_\nu\phi^c\,\,\,,
\ee
where the gauge field $A_{\mu}$ is defined through
\be
A_\mu=\frac{1}{\phi}\phi^a A^a_\mu \,\,\,.
\label{abelianfield}
\ee
\par
This solution has both electric and  magnetic fields. These
are given by
\bea
{\cal E}_i &=& {\cal F}_{i0} = \frac{x_i}{r}\frac{\rm d}{\rm d r}
\left(J(r)/er\right)\nonumber\\
{\cal B}_i&=&\frac{1}{2}\epsilon_{ijk}{\cal F}_{jk}=-\frac{1}{e}
\frac{x_i}{r^3}\,\,\,.
\eea
One can also calculate the electric and the magnetic charges
of this solution. In particular, the Dirac quantisation
condition gives the magnetic charge $g=1/e$.
\par
After this brief summary of the solutions to the $SO(3)$ gauge theory
let us express this solution in terms of the variables $G_{\mu\nu}^a$.
We will take for this purpose $c=-d=1$.  This choice leads
to $a=b=1$ and to the following
equation of motion for $G_{\mu\nu}^a$
\be
G^a_{\mu\nu}=-\frac{1}{2}\widetilde F^a_{\mu\nu}\,\,\,.
\ee
\par
In terms of the variables $G^a_{\mu\nu}$ an obvious gauge invariant
quantity is given by
\be
{\cal G}_{\mu\nu}= \frac{1}{\phi}\phi^a G_{\mu\nu}^a
+{1\over 2}\frac{1}{e\phi^3}\epsilon_{abc}
\epsilon_{\mu\nu\alpha\beta}
\phi^a D_\alpha\phi^b D_\beta\phi^c\,\,\,,
\ee
where the covariant derivative is in terms of
the gauge field as defined in (\ref{gaugefield}). The
$+{1\over 2}\epsilon_{\mu\nu\alpha\beta}$ tensor in the second
term is needed to get
${\cal G}_{\mu\nu}$ in a familiar form when expressed in terms of the
gauge filed $A_{\mu}$ as given in (\ref{abelianfield}). Indeed upon
using the equations of motion for $G_{\mu\nu}^a$
and the expression of $A_{\mu}$ we find that
\be
{\cal G}_{\mu\nu}=-{1\over 2}
\epsilon_{\mu\nu\alpha\beta}{\cal F}_{\alpha\beta}\,\,\,.
\ee
Hence if we identify ${\cal G}_{\mu\nu}$ as the electromagnetic field
in the dual theory, we see that what was previously called
the electric field  in the original theory
has become the magnetic field  in the dual theory
and vice-versa.
\par
In this paper we have found a way of constructing a classical
dual action for a general non-abelian and non-supersymmetric
gauge theory. The theory we obtained describes the dynamics
of a rank-two antisymmetric tensor field and could be studied
for its own right regardless of its origin.
\par
Going from the action $S$ in (\ref{action1}) to the action $I$ in
(\ref{action2}) can be made into a formal step in the path integral
by noticing that
\bea
&{}&\exp\left[\int {\rm d}^4x\left(
\tau_+\left(F^+\right)^2 + \tau_-\left(F^-\right)^2\right)\right]
\,\,\,\,\,\,\propto\nonumber\\
&{}&\int {\cal D}G^+{\cal D}G^-
\exp\left[\int {\rm d}^4x\left(
a\left(G^+\right)^2 + b\left(G^-\right)^2
+cF^+G^+ +d F^-G^-\right)\right]\,\,\,.
\eea
In this way one obtains a dual action at the quantum level and
the properties of the partition function could be examined
along the lines of refs.\cite{witten,verlinde}.
\newline
{\bf{Acknowledgements:}} I would like to thank Prof. Daniel Z. Freedman for
correspondence.
\newline
{\bf{Note Added:}}
I became aware, while this work was being proof-read, that
some related work has been done in ref.\cite{ganor} using
a different method and specialising to the
case of pure gauge theory
only.

\end{document}